\documentstyle[12pt]{article}

\begin{document}

\begin{center}
{\large {\bf COMMENTS ON \textquotedblleft NEW BRANS-DICKE WORMHOLES"}}\\

\vspace{8mm}

Arunava Bhadra,$^{a,}$\footnote{%
E-mail: aru\_bhadra@yahoo.com} Ion Simaciu,$^{d,}$\footnote{%
E-mail: isimaciu@yahoo.com} Kamal Kanti Nandi$^{c,e,}$\footnote{%
E-mail: kamalnandi@hotmail.com} and Yuan-Zhong Zhang$^{b,c,}$\footnote{%
E-mail: yzhang@itp.ac.cn}\\[0pt]
\vspace{5mm} {\footnotesize {\it \ $^a$ High Energy and Cosmic Ray Research
Center, University of North Bengal, Darjeeling (W.B.) \\[0pt]
734 430, India.\\[0pt]
$^b$ CCAST (World Laboratory), P.O.Box 8730, Beijing 100080, China.\\[0pt]
$^c$ Institute of Theoretical Physics, Chinese Academy of Sciences, P.O.Box
2735, Beijing 100080, China.\\[0pt]
$^d$ Department of Physics, University of ``Petrol-Gaze" Ploiesti, Ploiesti
2000, Romania.\\[0pt]
$^e$ Department of Mathematics, University of North Bengal, Darjeeling
(W.B.) 734 430, India. }}
\end{center}

\vspace{8mm}

\begin{abstract}
It is shown that the recently claimed two new Brans-Dicke wormhole
solutions [F. He and S-W. Kim, Phys. Rev. D{\bf 65}, 084022
(2002)] are not really new solutions. They are just the well known
Brans-Dicke solutions of Class I and II in a different conformal
gauge.

\bigskip

\noindent PACS number(s): 04.20.Gz, 04.50.+h
\end{abstract}

{\bf \vspace{20mm}}

There has been a revival of interest in the Brans-Dicke theory (BDT) in
recent times, particularly in the context of traversable Lorentzian
wormholes. Now a days, BDT is no longer regarded merely as a Machian
competitor to Einstein's General Relativity Theory (GRT) but a little more.
There are several reasons. The principal reason is of course that BDT
describes weak field tests of gravity reasonably well. Apart from this, it
is known that the BD scalar field $\phi $ plays the role of classical exotic
matter required for the construction of traversable Lorentzian wormholes [1].

In a recent paper, He and Kim [2] have found two new classes of solutions of
BDT and showed that the solutions represent massive Lorentzian traversable
wormholes. The purpose of this short Comment is to demonstrate that these
solutions can be alternatively derived by exploiting the conformal
invariance of the vacuum BDT action. That is, we show that the claimed
solutions are merely the Class I and II solutions of BDT in a different
gauge and thus are not essentially new.

We start from the BD action given by (we take units $G=c=1$):
$$
S_{BD}=\frac{1}{{16\pi }}\int {d^{4}x\sqrt{-g}\left[ {\phi R-\frac{\omega }{%
\phi }g^{\mu \nu }\nabla _{\mu }\phi \nabla _{\nu }\phi }\right] }+S_{matter}%
\eqno(1)
$$%
in which $S_{matter}$ represents the nongravitational part of the action
which is independent of the scalar field $\phi $. With Ref.[2], we consider
only the pure gravitational part of the action and set $S_{matter}=0$. Under
a conformal transformation, with a constant gauge parameter $\xi $,
$$
\tilde{g}_{\mu \nu }=\phi ^{2\xi }g\eqno(2)
$$%
the Lagrangian density becomes
$$
L_{BD}\sqrt{-g}=\sqrt{-\tilde{g}}\left[ {\phi ^{1-2\xi }\tilde{R}-6\phi
^{1-5\xi }\left( {\phi ^{\xi }}\right) _{;\mu }^{;\mu }-\frac{\omega }{{\phi
^{1+2\xi }}}\tilde{g}^{\mu \nu }\tilde{\nabla}_{\mu }\phi \tilde{\nabla}%
_{\nu }\phi }\right] .\eqno(3)
$$%
Under a further redefinition of the scalar field
$$
\tilde{\phi}=\phi ^{1-2\xi }\eqno(4)
$$%
the Lagrangian density simply becomes
$$
L_{BD}\sqrt{-g}=\sqrt{-\tilde{g}}\left[ {\tilde{\phi}\tilde{R}-\frac{{\tilde{%
\omega}}}{{\tilde{\phi}}}\tilde{g}^{\mu \nu }\tilde{\nabla}_{\mu }\tilde{\phi%
}\tilde{\nabla}_{\nu }\tilde{\phi}}\right] \eqno(5)
$$%
where
$$
\tilde{\omega}=\frac{{\omega -6\xi (\xi -1)}}{{(1-2\xi )^{2}}}.\eqno(6)
$$%
Therefore the vacuum BD action (1) is invariant under the transformations
(2) and (4). We briefly mention some relevant implications of this
invariance. Faraoni [3] has shown that the transformations (2) and (4) that
map $\left( {g_{\mu \nu },\phi }\right) \rightarrow \left( {\tilde{g}_{\mu
\nu },\tilde{\phi}}\right) $ constitute a one-parameter Abelian group with a
singularity in the parameter dependence at $\xi =1/2$. He used this
invariance to show that the $\omega \rightarrow \infty $ limit of the BDT
does not lead to vacuum GRT, when the matter stress energy (other than that
of $\phi $) is traceless. (However, for a critique of his arguments, see
Ref.[4]). Cho [5] called the above conformal invariance as indicating an
inherent ambiguity of the vacuum BD action and he argued that the only way
to resolve this ambiguity is to specify how the physical metric couples to
matter field. On the other hand, if one includes matter field in the action
(1), one ends up with \textquotedblleft abnormal" coupling with it in the
conformally rescaled action so that the principle of equivalence is violated
by the motion of ordinary matter. This violation is of no concern as there
are important gains: The so called ambiguity is removed and that the
gravitational interaction is described by spin-two massless graviton.

Let us introduce the parameter $\xi $ into the Class I solution of BDT [6]
described by the action (1). The general solution then looks like, using (2)
and (4),
$$
ds^{2}=-e^{2\alpha (r)}dt^{2}+e^{2\beta (r)}\left[ {dr^{2}+r^{2}d\theta
^{2}+r^{2}\sin ^{2}\theta d\varphi ^{2}}\right] \eqno(7)
$$%
where
$$
e^{\alpha (r)}=e^{\alpha _{0}}\left( {\frac{{1-\frac{B}{r}}}{{1+\frac{B}{r}}}%
}\right) ^{\frac{{1+\xi C}}{\lambda }}\eqno(8)
$$%
$$
e^{\beta (r)}=e^{\beta _{0}}\left( {1+\frac{B}{r}}\right) ^{2}\left( {\frac{{%
1-\frac{B}{r}}}{{1+\frac{B}{r}}}}\right) ^{\frac{{\lambda -C-1+\xi C}}{%
\lambda }}\eqno(9)
$$%
$$
\phi (r)=\phi _{0}\left( {\frac{{1-\frac{B}{r}}}{{1+\frac{B}{r}}}}\right) ^{%
\frac{{C(1-2\xi )}}{\lambda }}\eqno(10)
$$%
where the constants $\lambda ,C$ are still connected via the BD field
equations by
$$
\lambda ^{2}\equiv (C+1)^{2}-C\left( {1-\frac{{\omega C}}{2}}\right) .\eqno%
(11)
$$%
Evidently, the parameter $\xi $ has cancelled out and that is why we called
it a gauge:\ It can be fixed to any value. The arbitrary constants $\alpha
_{0}=0,\;\beta _{0}=0$ are determined by asymptotic flatness. Now choose the
gauge and redefine the constant $\lambda $ such that
$$
\xi =-\frac{1}{C},\quad \quad \frac{{C+2}}{\lambda }=C^{\prime }.\eqno(12)
$$%
Then the metric and the scalar function (8)-(10) immediately become
$$
e^{\alpha (r)}=1\eqno(13)
$$%
$$
e^{\beta (r)}=\left( {1+\frac{B}{r}}\right) ^{2}\left( {\frac{{1-\frac{B}{r}}%
}{{1+\frac{B}{r}}}}\right) ^{1-C^{\prime }}\eqno(14)
$$%
$$
\phi (r)=\phi _{0}\left( {\frac{{1-\frac{B}{r}}}{{1+\frac{B}{r}}}}\right)
^{C^{\prime }}.\eqno(15)
$$%
Using the Eqs.(12) and (6), the identity (11) can be rewritten as
$$
C^{\prime }{}^{2}=\left( {\frac{{\tilde{\omega}+2}}{2}}\right) ^{-1}.\eqno%
(16)
$$%
This is exactly the solution obtained in Ref.[2] on solving the complete set
of BD field equations {\it ab initio}. Similarly, consider the Class II BD
solution with the gauge $\xi $, viz.,
$$
\alpha (r)=\alpha _{0}+\frac{{2(1+\xi C)}}{\Lambda }\arctan \left( {\frac{r}{%
B}}\right) \eqno(17)
$$%
$$
\beta (r)=\beta _{0}-\frac{{2(C+1-\xi C)}}{\Lambda }\arctan \left( {\frac{r}{%
B}}\right) -\ln \left( {\frac{{r^{2}}}{{r^{2}+B^{2}}}}\right) \eqno(18)
$$%
$$
\phi =\phi _{0}e^{\left[ {\frac{{2C(1-2\xi )}}{\Lambda }}\right] \arctan
\left( {\frac{r}{B}}\right) }\eqno(19)
$$%
$$
\lambda ^{2}\equiv C\left( {1-\frac{{\omega C}}{2}}\right) -(C+1)^{2}.\eqno%
(20)
$$%
One can choose the gauge, again of the same form, $\xi =-\frac{1}{C},\quad
\frac{{C+2}}{\Lambda }=C^{\prime }$ to find that the resulting solution
precisely coincides with the other solution in Ref.[2]. The key point is
that, it is possible to generate an infinite set of formal solutions simply
by assigning arbitrary values to the gauge parameter $\xi $ and the
resulting sets of solutions form equivalence classes having the same
physical content as the original 1962 solutions.

It should be noted in passing that, under the identifications
$$
\tilde{\lambda}=\frac{\lambda }{{1+\xi C}},\quad \quad \tilde{C}=\frac{{%
C(1-2\xi )}}{{1+\xi C}},\eqno(21)
$$%
the solution set (8)-(11) has exactly the same form as the original BD Class
I solutions [6] with $\xi =0$. It is possible to obtain the solutions
(13)-(16) from the solution with $\xi =0$ simply by letting $C,\;\lambda
\rightarrow \infty $ such that $(C/\lambda )\rightarrow C^{\prime }$. This
is obviously equivalent to choosing the gauge $\xi =-1/C$ in Eqs.(21) so
that $\tilde{C},\;\tilde{\lambda}\rightarrow \infty $, but $(\tilde{C}/%
\tilde{\lambda})=C^{\prime }$, in virtue of Eq.(12). Similar considerations
apply for the solution set (17)-(20) in respect of BD Class II solutions.
The above analyses make it clear that conformal invariance of the vacuum BD
action can be used as a tool to separate genuinely new solutions from the
existing ones. This concludes what we wished to demonstrate.

\section*{Acknowledgments}

AB wishes to thank IRC-NBU for providing its journal facility. KKN
wishes to thank Professor Ou-Yang Zhong-Can for providing
hospitality and excellent working facilities at ITP, CAS.
Unstinted assistance from Sun Liqun is also gratefully
acknowledged. This work is supported in part by the TWAS-UNESCO
program of ICTP, Italy, in part by the financial assistance from
the Chinese Academy of Sciences as well as in part by the National
Basic Research Program of China under Grant No. 2003CB716300 and
by the NNSFC under Grant No. 90403032.

\end{document}